\documentclass[aps,prl,twocolumn,groupedaddress,showpacs,showkeys]{revtex4-1}

\usepackage{verbatim}
\usepackage{amsmath}
\usepackage{graphicx}
\usepackage{array}
\usepackage{SIunits}
\usepackage{color}
\usepackage{float}
\usepackage{placeins}
\usepackage{epstopdf}
\usepackage{upgreek}
\newcommand{\pryso}{Pr$^{3+}$:Y$_2$SiO$_5$}
\newcommand{\yso}{Y$_2$SiO$_5$}
\newcommand{\dhdh}{$^3\!H_4-\,^1D_2\,$}

\usepackage{hyperref}
\def\equationautorefname~#1\null{Eq. (#1)\null}
\def\figureautorefname~#1\null{Fig. #1\null}

\begin{document}


\title{Wave propagation in birefringent materials with off axis absorption or gain}


\author{Mahmood Sabooni$^{a,b}$}
\email[Email:]{m.sabooni@ut.ac.ir;mahmood.sabooni@fysik.lth.se}
\author{Adam N. Nilsson$^{b}$, Gerhard Kristensson$^{c}$, Lars Rippe$^{b}$}
\affiliation{$^{a}$ Department of Physics, University of Tehran, 14399-55961, Tehran, Iran}
\affiliation{$^{b}$ Department of Physics, Lund University, P.O.~Box 118, SE-22100 Lund, Sweden}
\affiliation{$^{c}$ Department of Electrical and Information Technology, Lund University, P.O.~Box 118, SE-22100 Lund, Sweden}


\date{\today}

\begin{abstract}
The polarization direction of an electromagnetic field changes and eventually reaches a steady state when propagating through a birefringent material with off axis absorption or gain. The steady state orientation direction depends on the magnitude of the absorption (gain) and the phase retardation rate. The change in the polarization direction is experimentally demonstrated in weakly doped ($0.05\%$) \pryso ~crystals, where the light polarization, if initially aligned along the most strongly absorbing principal axis, gradually switch to a much less absorbing polarization state during the propagation. This means that the absorption coefficient, $\alpha$, in birefringent materials generally varies with length. This is important for, e.g., laser crystal gain media, highly absorbing and narrow band spectral filters and quantum memories.
\end{abstract}


\maketitle


\section{Introduction\label{sec:Introduction}}
\par
For electromagnetic plane wave propagation in birefringent materials the instantaneous polarization direction of the electric field vector, $\mathbf{E}$, normally changes during the propagation. For example, the polarization of a wave propagating along one of the principal axes in a (non-absorbing) birefringent material with (initially) linear polarization at an angle $\gamma$ relative one of the other principal axes, changes from linear to elliptical (with major axes $|\mathbf{E}|\cos\gamma$ or $|\mathbf{E}|\sin\gamma$) and then to linear polarization at angle $-\gamma$ and then back to elliptical etcetera as it propagates through the material. For a case where the wave is not propagating along one of the principal axes the wave vector and the Poynting vector are in general not parallel and there are walk off effects (e.g. chapter 6 in Ref. \cite{Saleh2007}).

In this work we consider an initially linearly polarized plane wave, propagating along a principal axis in a non-magnetic off axis absorbing birefringent material. In this case polarization rotation can occur even if the initial linear polarization is aligned with one of the principal axes. In materials with absorption or gain the development of the polarization can be complicated. Not only could there be oscillations between linear and elliptical polarization and walk off effects, there can also be conversions from light of one polarization to another polarization due to the absorption (or gain).

Such effects should for example be present in laser crystal gain media when the gain is anisotropic and the gain tensor is not aligned with the principal axes \cite{Digonnet2001}. Recently there has also been an interest in highly absorbing and very narrow bandwidth filters for very specific applications such as quantum memories \cite{Riedmatten2015}, or highly absorbing narrowband filters with exceptionally large etendue for high performance ultrasound optical tomography \cite{Louchet-Chauvet2011,Zhang2012} and also dynamically tunable high performance filters \cite{Beavan2013a}. All these recent papers use filters based on rare-earth-ion-doped inorganic crystals where the absorption tensor is not aligned with the principal axes. As a consequence the light polarization gradually switches during propagation to a much less absorbing polarization state even if the input polarization is perfectly aligned with the most strongly absorbing principal axis. This significantly decreases the achievable filter attenuation from what might be anticipated based on just the material absorption coefficient.

In this work we theoretically analyze the polarization rotation in absorbing birefringent materials with a model adapted from \cite{Rikte2001}. We provide some simple relations of when the effects need to be taken into account and give possible suggestions on how to maintain the original high absorption. The theoretical analysis is supported by an experimental demonstration of the polarization rotation effect.

\par
The paper is organized as follows. First an intuitive theoretical background is given. This is followed by simulation results and discussions of polarization steady states and the effects of absorption on light propagating through a crystal. Experimental results are then presented where the incoming and outgoing polarization directions are studied for crystals with different absorption and length. Lastly a remark about the maximum absorption axis is made before the paper is concluded with a summary. A rigorous mathematical treatment of the problem can be found in the Supplementary Information section.

\section{Theoretical background}\label{sec:Theory}
\par
\begin{figure}[ht]
    \includegraphics[width=9cm]{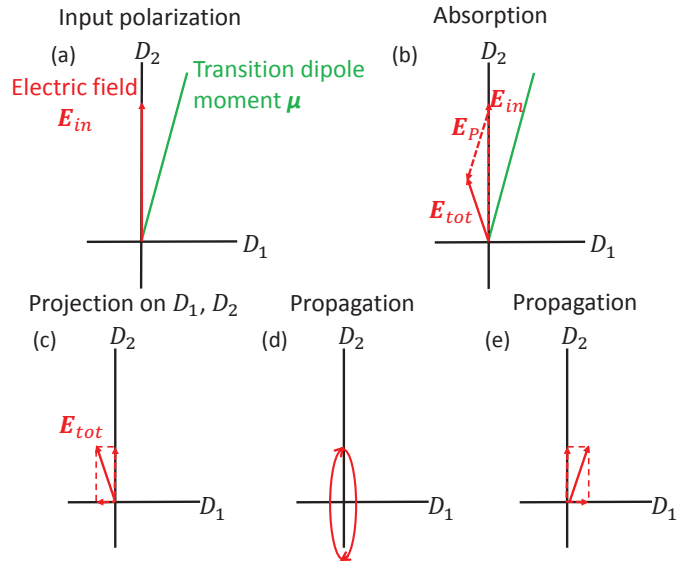}
    \caption{(Color online) Simplified view of the propagation of an electric field through a birefringent crystal with a transition dipole moment axis tilted with respect to the crystal axes ($D_1\bot D_2$). (a) The incoming electric field, $\mathbf{E}_{in}$, is polarized along $D_2$. (b) The material is polarized and emits an electric field $\mathbf{E}_P$ along the transition dipole moment which is $180\degree$ out of phase with the incoming field $\mathbf{E}_{in}$ resulting in the total field $\mathbf{E}_{tot} = \mathbf{E}_{in}+\mathbf{E}_P$ which has both a $D_1$ and $D_2$ component of the electric field. The magnitude of $\mathbf{E}_P$ is exaggerated to make the effects easier to see. The electric field is projected along the crystal axes $D_1$ and $D_2$ in (c) to make it possible to see the effect of propagation through the birefringent crystal which causes a relative phase change between the two electric field components. In (c) they are oscillating in phase, while in (d) they are $90\degree$ out of phase and in (e) they are completely out of phase ($180\degree$). }
    \label{fig:polRotExplain}
\end{figure}

A detailed description of the theoretical framework is given in the Supplementary Information, but a simple and intuitive understanding can be achieved by considering the effects of birefringence and absorption separately. In \autoref{fig:polRotExplain}a we see the principal axes of a birefringent crystal as well as the direction of the transition dipole moment, $\boldsymbol\mu$. This direction does not coincide with any of the principal axes of the crystal. An incoming field, $\mathbf{E}_{in}$, which is on resonance with the absorbing transition and with its polarization oriented along one of the principal axes, $D_2$, generates a polarization, $\mathbf{P}$, along $\boldsymbol\mu$ with a $90\degree$ phase shift. This polarization generates a field, $\mathbf{E}_P$, along $\boldsymbol\mu$ with a relative phase relation to $\mathbf{E}_{in}$ as shown in \autoref{fig:polRotExplain}b. The total resulting electromagnetic field, $\mathbf{E}_{tot} = \mathbf{E}_{in}+\mathbf{E}_P$, now also have a component along the $D_1$ axis as shown in \autoref{fig:polRotExplain}c. Since the crystal is birefringent the polarization state oscillates between linear (\autoref{fig:polRotExplain}c and \autoref{fig:polRotExplain}e) and elliptical (\autoref{fig:polRotExplain}d) as the field propagates through the crystal. Of course in reality the effects of absorption and birefringence are intertwined and occur simultaneously. The theoretical examination in the Supplementary Information makes it possible to simulate the propagation of any incoming electric field polarization through a birefringent crystal with an absorption axis with an angle to the principal axes. The method works for infinite incoming plane waves and assumes that the lateral (transverse) dimensions of the crystal is infinite as well. Discussions around the results from these simulations can be found in the next section.

\section{Simulation results}\label{sec:Simulation}
\par
In this section results from simulations of light propagation inside a birefringent crystal with a tilted transition dipole moment direction is analyzed. As a specific example a $0.05\%$ \pryso ~crystal is chosen where the transition dipole moment is tilted by $74.6 \pm 1.9\degree$ from the $D_1$ crystal axis \cite{GuokuiLiu2005} as seen in \autoref{fig:dipoleMoment} and calculated in the Supplementary Information, see \autoref{eq:chi_absorption2}.
\begin{figure}[ht]
    \includegraphics[width=8cm]{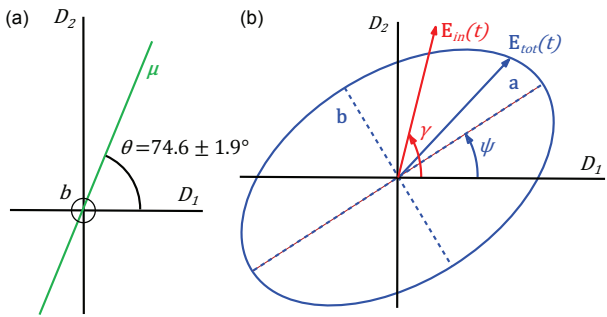}
    \caption{(Color online) (a) Transition dipole moment, $\boldsymbol\mu$, versus principal axes for \pryso ($D_1\bot D_2 \bot b$). (b) Notation of linear incoming and elliptical outgoing polarizations are seen where $\gamma$ is the angle between the $D_1$ axis and the incoming polarization and $\psi$ is the angle between the $D_1$ axis and the major axis, \textbf{a}, of the outgoing elliptical polarization.}
    \label{fig:dipoleMoment}
\end{figure}

\subsection{Phase retardation in a birefringent crystal}\label{sec:Phase}
The propagation phase, $\phi$, can be written as $\phi=2\pi \frac{nz}{\lambda_0}$ radians, i.e. light with the wavelength $\lambda_0$ in vacuum experiences a phase shift of $\phi$ radians when traveling an optical length $nz$ through a material, where $n$ is the index of refraction and $z$ is the propagation depth. The difference between the accumulated phases, $\phi_2$ and $\phi_1$, for light propagating along two different crystal axes, $D_2$ and $D_1$ respectively, is termed the phase retardation;
\begin{equation}{\label{eq:Retardation}}
\Delta\phi=\phi_2 - \phi_1 = 2\pi\frac{\Delta nz}{\lambda_0}
\end{equation}
where $\Delta n=n_{D_2}-n_{D_1}$. The crystal length needed to obtain a quarter-wave plate, $\Delta\phi=\frac{\pi}{2}$, for a \yso ~crystal with the different refractive indices; $n_{D_1}=1.7881$,  $n_{D_2}=1.809$ and $n_b=1.7851$ along the principal axes \cite{Beach1990} is calculated using \autoref{eq:Retardation} to be about $7.25\,\upmu$m for light propagating along the $b$ axis.

\par
\subsection{Polarization steady states}
From the effects explained in \autoref{fig:polRotExplain} it is clear that it is not possible to maintain a pure linear polarization when propagating through a crystal where the transition dipole moment axis does not coincide with any of the optical axes of the crystal. However, in general two steady state polarization solutions exist for a forward propagating wave. They are elliptically polarized and only differ by a $90\degree$ rotation of their major axes in the $D_1$-$D_2$ plane.  The polarization of the steady state does not change when propagating through the crystal. The ellipticity and the direction of these polarizations depend on the following ratio;

\begin{equation} {\label{eq:ratio}}
	R = \frac{n_{D_2}^2-n_{D_1}^2}{\chi_{abs}}
\end{equation}
where $\chi_{abs}$ is the electric susceptibility connected to the absorption (for \pryso ~see \autoref{eq:suscept_permitt} in the Supplementary Information). In general only the steady state solution with the lowest absorption or highest gain is stable.

\begin{figure*}[ht]
    \includegraphics[width=18cm]{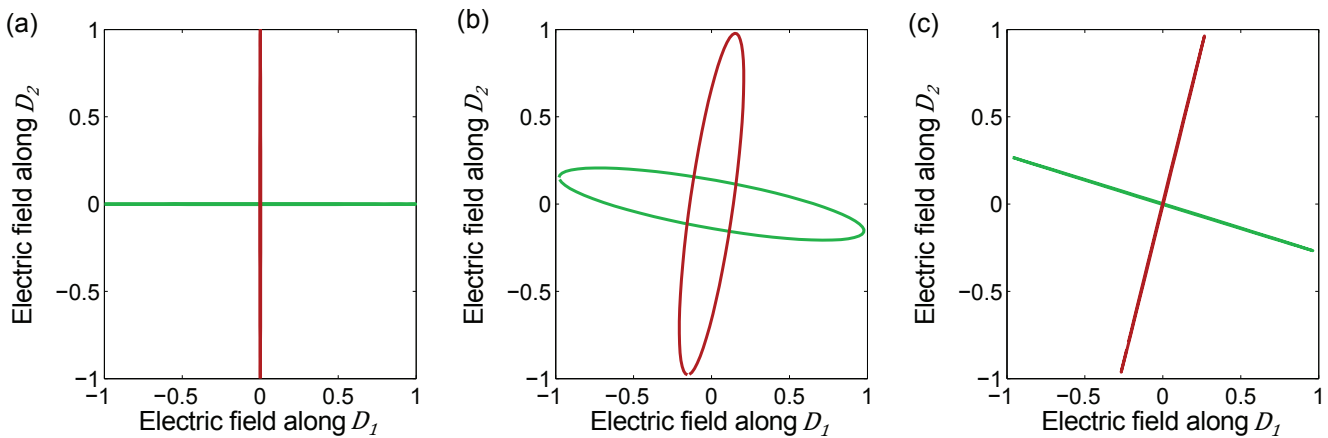}
    \caption{(Color online) Polarization steady states for (a) \pryso ~with absorption coefficients $\alpha_{D_1} = 3.6 \pm 0.5 \text{ cm}^{-1}$ and $\alpha_{D_2} = 47 \pm 5 \text{ cm}^{-1}$ which corresponds to an absorption susceptibility $\chi_{abs}=(8.82 \pm 0.8) \cdot10^{-4}$ and a transition dipole moment direction with an angle $\theta=74.6 \pm 1.9\degree$ from $D_1$ as can be seen in \autoref{fig:dipoleMoment}a. (b) and (c) has an increased absorption susceptibility, $\chi_{abs}$, by a factor of $100$ and $10 \text{ } 000$ respectively. The ratio $R$ in \autoref{eq:ratio} is (a) $R\approx 85$, (b) $R\approx 0.85$ and (c) $R\approx 0.0085$. The green curves are stable (unstable) and the red curves are unstable (stable) in the case of absorption (gain). }
    \label{fig:steadyStates}
\end{figure*}

In \autoref{fig:steadyStates}a-c the steady state polarization solutions when the transition dipole moment makes an angle of $74.6\degree$ relative to the $D_1$ axis are shown for three different cases where $R \gg 1$, $R \approx 1$ and $R \ll 1$, respectively. For absorption the green curves are stable solutions and the red curves are unstable. The reverse is true in the case of gain. The first case, seen in \autoref{fig:steadyStates}a, is the same as for a $0.05\%$ \pryso ~crystal which has a transition dipole moment direction that is tilted $74.6 \pm 1.9\degree$ from $D_1$ and absorption coefficients along $D_1$ and $D_2$ of $\alpha_{D_1} = 3.6 \pm 0.5 \text{ cm}^{-1}$ and $\alpha_{D_2} = 47 \pm 5 \text{ cm}^{-1}$ which corresponds to an electric susceptibility for the absorption that is $\chi_{abs}=(8.82 \pm 0.8)\cdot10^{-4}$. The difference $n_{D_2}^2-n_{D_1}^2$ in \yso ~is $0.075$ which gives a ratio $R \approx 85 \gg 1$. Here one can see that the steady state solutions lie respectively along the $D_1$ and the $D_2$ crystal axes, which is always the case for $R \gg 1$.

The next two cases seen in \autoref{fig:steadyStates}b and \autoref{fig:steadyStates}c use the same values mentioned above except that the absorption susceptibility, $\chi_{abs}$, is increased by a factor $100$ and $10 \text{ } 000$ respectively. In the extreme where $R \ll 1$, seen in \autoref{fig:steadyStates}c, the solutions lies almost completely along the transition dipole moment (unstable for absorption) and perpendicular to it (stable for absorption). In between these extreme cases when $R \approx 1$, seen in \autoref{fig:steadyStates}b, the solutions lie between the crystal axes and the transition dipole moments and have a higher ellipticity than in the extremes.

The steady state solutions can be explained by the same phenomena that was discussed in \autoref{fig:polRotExplain}, i.e. that an electric field in $D_2$ creates an electric field in $D_1$, and vice versa, due to the tilted absorption (gain) axis. The steady state solutions are polarization states where the sum of the components lost and absorbed in $D_1$ decays at the same rate as the sum of the components created and absorbed along $D_2$, which in other words means that the polarization stays the same.

The dependence on $R$ can be intuitively explained by rewriting the numerator in \autoref{eq:ratio} as;

\begin{align}{\label{eq:numerator_ratio}}
	n_{D_2}^2-n_{D_1}^2 &= (n_{D_2}+n_{D_1})(n_{D_2}-n_{D_1}) \nonumber\\
  	&= (n_{D_1}+n_{D_1}) \Delta n
\end{align}

If we now assume that $n_{D_1}$ and $n_{D_2}$ are roughly constant then the ratio $R$ measures if $\Delta n$ is small/large compared to the absorption and $\Delta n$ sets the rate at which phase retardation is accumulated (see \autoref{eq:Retardation}). In other words if $R \gg 1$ then the phase retardation dominates over the absorption. This makes all polarizations that are symmetric along the crystal axes equivalent (due to the effects explained in \autoref{fig:dipoleMoment}c-e) and therefore the only steady state solutions that can exist are along $D_1$ and $D_2$, see \autoref{fig:steadyStates}a. In the other extreme where $R \ll 1$ the absorption dominates and any phase retardation between the two electric field components can be ignored which leads to one solution orthogonal and one solution parallel to the transition dipole moment axis, see \autoref{fig:steadyStates}c.

\par
\subsection{Electric field absorption when $R \gg 1$}
\begin{figure*}[ht]
    \includegraphics[width=18cm]{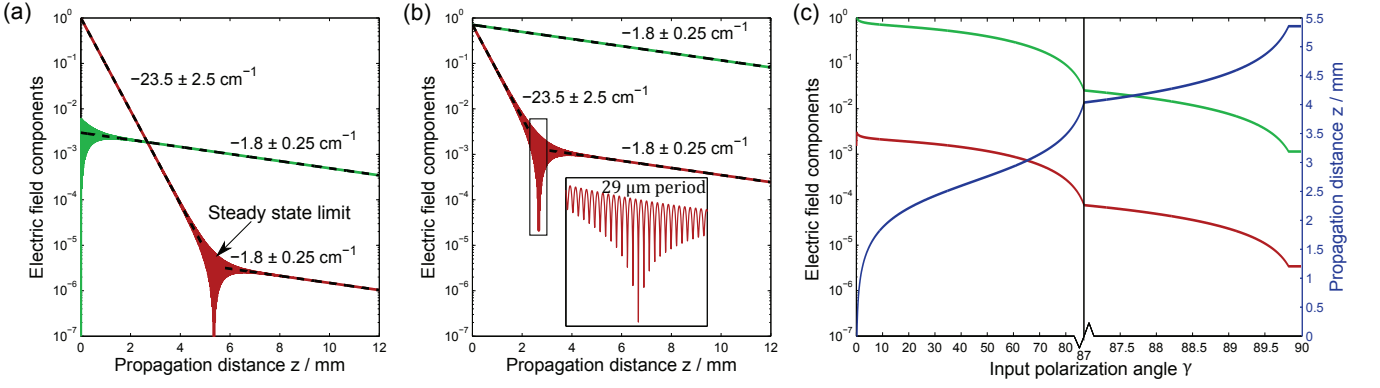}
    \caption{(Color online) Shows the decay of the electric field components along crystal axes $D_1$ (green) and $D_2$ (red) as a function of propagation distance $z$ for a linear input polarization at an angle (a) $\gamma = 90\degree$ and (b) $\gamma = 45\degree$ from $D_1$. The dashed black lines are linear fits to the logarithm of the electric field on the form $\text{ln}(E(z)) = \text{ln}(E_0) + C\cdot z$, where the value of $C$ is displayed next to each black curve. Note that these values represent the decay of the electric field. To obtain the absorption coefficients for the intensity multiply these values with $2$, i.e. $\alpha_{D_1} = 3.6 \pm 0.5\text{ cm}^{-1}$ and $\alpha_{D_2} = 47 \pm 5 \text{ cm}^{-1}$. The steady state limit can be seen in (a) which shows the largest total absorption that can be reached with a high absorption rate for a linear polarization. The inset in (b) is a zoomed in view of the steady state limit reached after $\approx 2.7$ mm for $\gamma=45\degree$. The oscillating pattern has a period of $29$ $\upmu$m (matching a $2\pi$ phase evolution calculated with \autoref{eq:Retardation}) and exists due to the interactions between the polarization induced in the material and the electric field of the light and the birefringence of the crystal. (c) The (approximate) propagation distance needed to reach the steady state limit shown in blue as a function of the input polarization angle $\gamma$ together with the electric field component along $D_1$ (green) and along $D_2$ (red) at the steady state limit. Note that the $x$-axis is cut at $87 \degree$ and that the axis scale changes afterwards. Note also the plateau reached at $89.83 \degree$ which shows that better angle precision than that is not needed. }
    \label{fig:decay}
\end{figure*}

From the simulations one can obtain the (expected) exponential decay in the electric field components along $D_1$ and $D_2$ as a function of propagation. For the example of \pryso ~the decay coefficients can be seen in \autoref{fig:decay}a for a linear input polarization along $D_2$ ($\gamma = 90\degree$ in \autoref{fig:dipoleMoment}b). Even though the input polarization only has a $D_2$ component of the electric field, a $D_1$ component is almost immediately created in the crystal due to the effects explained in \autoref{fig:polRotExplain}a-b. The relative size of the created component is the same as the ratio of the $D_2$ and $D_1$ component of the steady state, since at the steady state (when the polarization does not any longer change with propagation) the $D_2$ component is almost completely created by the tilted absorption of the $D_1$ component. Therefore it is a measure of how strong the connection between the two components is. For the case of \pryso ~the initial electric field component along $D_1$ is approximately a factor of $330$ less than the $D_2$ component (see \autoref{fig:decay}a and \autoref{eq:ratio_electric2} in the Supplementary Information).

The exponential decay along $D_2$ is much larger than along $D_1$ in the beginning of the propagation due to the fact that the transition dipole moment lies much closer to $D_2$. However towards the end of the propagation the stable steady state solution is reached and both components decay at the same rate. Continuing in \autoref{fig:decay}b the input polarization is changed to $\gamma = 45 \degree$ and the polarization steady state is reached sooner since the input polarization is more similar to the stable steady state. For a linear input polarization the highest absorption rate is achieved along $D_2$ but slows down after propagating around $5.4$ mm, see \autoref{fig:decay}a. This leaves us with the ultimate limit, which we call the ``steady state limit'', of the maximum electric field absorption with a high absorption rate of the incoming light for this material to be $10^{-6}-10^{-5}$ and this is reached for a $5.4$ mm long crystal. To clarify, the light continues to be absorbed after this limit but at a much slower rate. In light of this limit it is clear that scaling a crystal longer than $\approx 5.4$ mm to gain absorption serves limited purpose and one can get exponentially higher absorption by avoiding the steady state limit. Note also that much earlier, at around $2.7$ mm, the electric field component along $D_1$, which decays much slower, is of the same size as the component along $D_2$ and the $D_1$ component dominates after $2.7$ mm, even though the $D_1$ component was zero before entering the crystal, this component can however be blocked with a polarizer after the crystal.

The (approximate) propagation distance needed to reach the steady state limit as a function of the input polarization $\gamma$ is shown in blue in \autoref{fig:decay}c. The electric field component along $D_1$ and $D_2$ when the steady state limit is reached is also shown in green and red respectively. Note that the $D_1$ electric field component is always a factor of $\approx330$ larger than the $D_2$ component when the steady state solution is reached. Close to $\gamma = 90\degree$ the propagation distance decreases quite rapidly as $\gamma$ departs from $90\degree$ and a good control over the purity of the electric field polarization is required to have a high absorption and a steady state limit at $5.4$ mm. The purity needed is in the order of the ratio of the $D_1$ and $D_2$ electric field components for the stable steady state, which for \pryso, seen in \autoref{fig:steadyStates}a, is a factor $\approx330$ for the electric field, or equivalently a factor $10^5$ for the intensity components (see \autoref{eq:ratio_electric2} in the Supplementary Information). This means that a polarizer with better suppression than $10^5$ and an angle precision of around $0.17 \degree$ is required to achieve the longest propagation distance with maximum absorption.

To avoid the steady state limit one has to suppress the $D_1$ component of the light. This can be done with e.g. a linear polarizer. To achieve high absorption it is therefore better to use several smaller crystals (lengths of $\leq 5.4$ mm) with linear polarizers blocking any $D_1$ component in between each set of crystals instead of using one long crystal.

\par
\section{Experimental results}
\par
\begin{figure}[ht]
    \includegraphics[width=9cm]{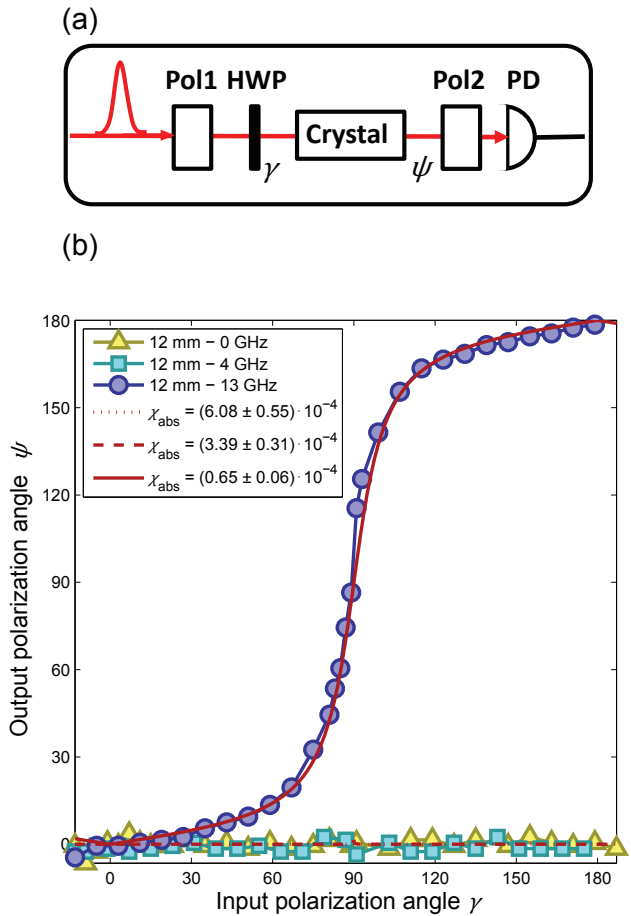}
    \caption{(Color online) (a) Polarization analyzer setup. First, two polarizers are calibrated against each other by minimizing the photodiode (PD) signal when there is nothing in between them. The first polarizer (Pol1) and the half-wave plate (HWP) makes sure that the incoming linearly polarized light is at an angle $\gamma$ from $D_1$. The $D_1$ and $D_2$ directions are measured with the same setup with the crystal at room temperature where absorption is negligible. By turning the second polarizer (Pol2) to maximize the throughput to the photodiode (PD) the outgoing polarization after the crystal is measured to be $\psi$ . (b) Three experiments were performed on a $12$ mm crystal $0$, $4$ and $13$ GHz away from the center of the inhomogeneous absorption profile. The simulations have $\chi_{abs}$ given by \autoref{eq:lorentz} which assumes an inhomogeneous profile in the shape of a Lorentzian distribution using a FWHM of $9$ GHz, which was measured in the experiment. The experimental results are displayed as yellow triangles, cyan squares and blue dots together with a red dotted, dashed and solid curve respectively for the three cases of $0$, $4$ and $13$ GHz. Note that the simulation results for the $0$ and $4$ GHz cases overlap.}
     \label{fig:polarization_rotation}
\end{figure}

To examine the polarization change during propagation several experiments were performed on crystals with different lengths. We used three 0.05\% \pryso ~crystals with different lengths ($1$ mm, $6$ mm and $12$ mm) and measured the outgoing polarization direction $\psi$ as a function of the incoming polarization angle $\gamma$, see \autoref{fig:dipoleMoment}b. The crystals were kept at cryogenic temperature ($2$ K). The setup is illustrated in \autoref{fig:polarization_rotation}a. The absorption was probed with attenuated pulses to minimize saturation effects on the absorption and between each experiment a set of hole eraser pulses were used over a wide spectral range to make sure that no permanent hole burning occurred.

According to the simulations for a \pryso ~crystal there is a minimum (sufficient) value of the absorption where the steady state solution is reached which means that the outgoing polarization is almost entirely along the $D_1$ axis. A $12$ mm \pryso ~crystal is employed to illustrate this effect as shown in \autoref{fig:polarization_rotation}b, where three measurements at different frequencies within the inhomogeneous profile were performed. The inhomogeneous line-width, $\Gamma$, of this crystal is measured to be about $9$ GHz. The three sets of measurements at frequencies $0$ GHz, $4$ GHz and $13$ GHz relative to the line center are shown in \autoref{fig:polarization_rotation}b as yellow triangles, cyan squares and blue dots respectively together with simulations for each case displayed as a red dotted, dashed and solid curve. For high and intermediate absorption the outgoing polarization is along the $D_1$ axis which agrees with the fact that the $D_2$ component is absorbed much quicker than the $D_1$ component.

As shown in \autoref{fig:polRotExplain}c-e the outgoing polarization shifts significantly during propagation. It is therefore impossible to determine if the outgoing polarization is in the range $\psi=0-90\degree$ or $\psi=90-180\degree$ for a given input polarization without a very precise knowledge of the crystal length. A quarter-wave plate distance was calculated previously to be $L\approx 7.25$ $\upmu$m which means that a full evolution of the phase is done in $\approx 29$ $\upmu$m in \yso. If the crystal length is not known within this precision it is indeed impossible to determine the outgoing polarization direction. The crystal length used in the simulations is therefore changed from exactly $12$ mm by a few tens of $\upmu$m to the value that best matches the experimental results.

The absorption coefficients in \cite{GuokuiLiu2005} are measured in a crystal with a line-width of $\Gamma = 6.2$ GHz and which nominally has the same Pr concentration, $0.05 \%$ as in the present crystal. This would correspond to a peak absorption susceptibility of $\chi_{abs}^{peak} = (8.82 \pm 0.8) \cdot 10^{-4}$. The inhomogeneous profile can be approximated as a Lorentzian profile \cite{GuokuiLiu2005} that in our experiments have a line-width of $\Gamma=9$ GHz. Assuming that the integral of the absorption should be the same for our crystal and the crystal used in \cite{GuokuiLiu2005} the peak absorption susceptibility of our crystal should be reduced by a factor $9/6.2$ which gives $\chi_{abs}^{peak} = (6.08 \pm 0.55) \cdot 10^{-4}$. Since the three measurements seen in \autoref{fig:polarization_rotation}b are performed at different positions on the inhomogeneous profile the absorption susceptibility used in the simulations must be adjusted accordingly;
\begin{equation} {\label{eq:lorentz}}
\begin{cases}
	\chi_{abs} = \chi_{abs}^{peak} \frac{(\Gamma/2)^2}{(f-f_0)^2 + (\Gamma/2)^2} \\
	\chi_{abs,1} = [f-f_0=0 \text{ GHz}] \approx   (6.08\pm0.55) \cdot 10^{-4}	 \\
	\chi_{abs,2} = [f-f_0=4 \text{ GHz}] \approx   (3.39\pm0.31) \cdot 10^{-4}\\
	\chi_{abs,3} = [f-f_0=13 \text{ GHz}] \approx (0.65\pm0.06) \cdot 10^{-4}
\end{cases}
\end{equation}

The simulations agree well with the experimental results and only in the $13$ GHz case is the absorption so low that the outgoing polarization varies from $D_1$. The results from the $1$ and $6$ mm crystals give similar results that agree with the simulations. Also here the absorption susceptibility must be reduced with respect to the peak value to account for the experiments taking place at the side of the inhomogeneous profile.

\par
\section{Maximum absorption axis}
\par
It has previously been shown, see Ref. \cite{Walther2009a}, that slow light effects split an incoming linear polarization into two parts, where one part has a significantly larger slow light effect than the other. The transition dipole moment direction was then considered as the maximum absorption direction but it turns out that the maximum (minimum) absorption and consequently time delay happens when we set the input polarization along the $D_2$ ($D_1$) axis. In general this depends on the ratio $R$ in \autoref{eq:ratio}, since if $R \ll 1$ the maximum absorption direction is almost exactly along the transition dipole moment axis, but as stated this is not the case for \pryso ~where $R \approx 85 \gg 1$ .
\par

\section{Conclusion}
\par
In conclusion, we have modeled and analyzed the polarization of light propagating through an absorbing (amplifying) birefringent crystal. Steady state solutions for the polarization of a propagating wave were found and discussed. This led to the conclusion that only increasing the length of a birefringent medium (e.g. a rare-earth-ion-doped crystal) is not the most efficient method to reach higher absorption. This can be an important issue in applications like quantum memories and spectral filtering employing birefringent crystals \cite{Riedmatten2015,Louchet-Chauvet2011,Zhang2012,Beavan2013a}. Since the model works for any birefringent material with either absorption or amplification it can be useful in other applications such as laser crystal gain media or with other materials that are not specifically discussed here.

It also became clear that the polarization direction with maximum absorption (gain) depends upon the ratio between the phase retardation and the magnitude of the absorption (gain). For an absorbing \pryso ~crystal the maximum absorption axis is along $D_2$ while the stable polarization steady state solution is almost completely along $D_1$.

This work will hopefully open up new opportunities to investigate more interesting physics and applications regarding the propagation effects in rare-earth-ion-doped crystals in the future.

\par
This work was supported by the Swedish Research Council, the Knut \& Alice Wallenberg Foundation, the Crafoord Foundation, the EC FP7 Contract No. 247743 (QuRep). The research leading to these results also received funding from the People Programme (Marie Curie Actions) of the European Union's Seventh Framework Programme FP7 (2007-2013) under REA grant agreement no. 287252(Marie Curie Action). Finally, we are most grateful to Prof. Stefan Kr\"{o}ll and Dr. Mikael Afzelius for several valuable discussions.\\

\bibliography{C:/Users/Mahmood/Dropbox/PAPERS/Main_Ref}

\clearpage
\section{Supplementary information}
\subsection{Light propagation in a birefringent crystal}
\par
The propagation of an electromagnetic field in a general material containing absorption can be a quite complex problem to solve due to the interaction of light and matter, but one could reduce this complexity without loosing the generality of the solution by considering a simplification based on the properties of the medium. We start the simplification by considering a plane-stratified medium, which is a valid assumption for the materials investigated in our case. The assumption of plane-stratified media is a proper approximation for materials that show variation in the propagation direction ($z$-axis) but no variation in the lateral ($x$-$y$ plane) direction. In other words, we have homogeneity (same absorption or gain) in the $x$-$y$ plane and possible inhomogeneity (varying absorption or gain) along the $z$-axis. The main theoretical concept of the calculations in this paper is discussed with more details in Ref. \cite{Rikte2001}.

\par
For macroscopic media, Maxwell's equations describe the dynamics of the fields as follows (the harmonic time convention $e^{-i\omega t}$ is used):
\begin{equation}{\label{eq:Maxwell}}
\begin{cases}
  \bigtriangledown\times \mathbf{E(r,\omega)} = ik_0(c_0\mathbf{B(r,\omega)})\\
  \bigtriangledown\times \eta_0\mathbf{H(r,\omega)} = -ik_0(c_0\eta_0\mathbf{D(r,\omega)})\\
\end{cases}
\end{equation}
where $\eta_0=\sqrt[]{\frac{\mu_0}{\epsilon_0}}$ is the intrinsic impedance of vacuum, $c_0=1/\sqrt[]{\epsilon_0\mu_0}$ and $k_0$ are the speed and wave number of light in vacuum, respectively. In addition, \textbf{E}, \textbf{B}, $\textbf{H}$, and $\textbf{D}$ are the electric field, magnetic flux density, magnetic field and the electric flux density, respectively.

In order to apply Maxwell's macroscopic equations, it is necessary to specify the relations between $\textbf{E}$, $\textbf{B}$, $\mathbf{H}$, and $\mathbf{D}$. These equations are called the \emph{constitutive relations}.
\begin{equation}{\label{eq:constitutive}}
\begin{cases}
  \mathbf{D(r, \omega)} = \epsilon_0[\boldsymbol{\epsilon}(z,\omega)\cdot\mathbf{E(r,\omega)}+\eta_0\boldsymbol{\xi}(z,\omega)\cdot\mathbf{H(r,\omega)}]\\
  \mathbf{B(r, \omega)} = \frac{1}{c_0}[\boldsymbol{\zeta}(z,\omega)\cdot\mathbf{E(r,\omega)}+\eta_0\boldsymbol{\mu}(z,\omega)\cdot\mathbf{H(r,\omega)}]\\
\end{cases}
\end{equation}
where $\mathbf{r}=x\mathbf{\hat{x}}+y\mathbf{\hat{y}}+z\mathbf{\hat{z}}$ and $\omega$ is the angular frequency. $\boldsymbol\epsilon$ and $\boldsymbol\mu$ are the \emph{permittivity} dyadic and \emph{permeability} dyadic of the medium, respectively, while $\boldsymbol\xi$ and $\boldsymbol\zeta$ are called crossed magneto-electric dyadics.  All four dyadics and all four fields can depend on the angular frequency $\omega$, but this variable is suppressed below to simplify the notation.
\par
In a single crystal, the physical and mechanical properties can often be orientation dependent. When the properties of a material vary with orientation, the material is said to be \emph{anisotropic}. Alternatively, when the properties of a material are the same in all directions, the material is said to be \emph{isotropic}. \emph{Bi-anisotropic} is a general class of linear media which exhibit so-called magnetoelectric coupling between the electric and magnetic fields \cite{Rikte2001}.

\par
Based on the plane-stratified assumption, it is natural to decompose the electromagnetic field into tangential ($x$-$y$ plane) and normal components. Considering the lateral homogeneity and substituting the constitutive relations (\autoref{eq:constitutive}) into Maxwell's equations (\autoref{eq:Maxwell}) gives a system of ordinary differential equations (ODEs) with the variable $z$.

\begin{equation}{\label{eq:wave_eq}}
\frac{d}{dz}
\begin{pmatrix}
  \mathbf{E}_{xy}(z)\\
  \eta_0\mathbf{J}\cdot\mathbf{H}_{xy}(z)\\
  \end{pmatrix}
  = ik_0\mathbf{M}(z)\cdot
  \begin{pmatrix}
  \mathbf{E}_{xy}(z)\\
  \eta_0\mathbf{J}\cdot\mathbf{H}_{xy}(z)\\
  \end{pmatrix}
\end{equation}
where $\mathbf{J}$ is a two-dimensional rotation dyadic (rotation of $90\degree$ in the $x$-$y$ plane) as follows:
\begin{equation}{\label{eq:J}}
\mathbf{J}=
\begin{pmatrix}
  0 & -1\\
  1 & 0\\
\end{pmatrix}
\end{equation}
and
\begin{equation}{\label{eq:I}}
\mathbf{J}\cdot\mathbf{J}=-\mathbf{I_2}=
\begin{pmatrix}
  -1 & 0\\
  0 & -1\\
\end{pmatrix}
\end{equation}

For a more detailed discussion of how to obtain \autoref{eq:wave_eq}, which is beyond the scope of this paper, the reader is referred to Ref. \cite{Rikte2001}.

According to the main assumption, the plane-stratified media assumption, the four dyadics $\boldsymbol{\epsilon}(z),\boldsymbol{\xi}(z),\boldsymbol{\zeta}(z),$ and $\boldsymbol{\mu}(z)$ in \autoref{eq:constitutive} depend only on one spatial variable, $z$, which describes the propagation distance. The \pryso~crystal, which is our medium, is appropriately modeled as a non-magnetic anisotropic medium with zero crossed magneto-electric dyadics. Therefore, as a second assumption, we assume non-magnetic properties ($\boldsymbol{\mu}=\mathbf{I}$) and also no crossed magneto-electric dyadics ($\boldsymbol{\xi}=\boldsymbol{\zeta}=0$) in \autoref{eq:constitutive}.  There exists a general set of equations for the connection between the fundamental matrix $\mathbf{M}(z)$ and the four main dyadics in Ref. \cite{Rikte2001}, but based on the three assumptions described above and a normal incident we can simplify $\mathbf{M}(z)=\mathbf{M}$ for our case as:

\begin{equation}{\label{eq:Melements}}
\mathbf{M}=
\begin{pmatrix}
  \mathbf{\bar{\bar{0}}} & -\mathbf{\bar{\bar{I}}}_2\\
  -\bar{\bar{\boldsymbol\epsilon}}_{\perp\perp}+\frac{1}{\epsilon_{zz}} \bar{\boldsymbol\epsilon}_{\perp} \bar{\boldsymbol\epsilon}_{z} & \mathbf{\bar{\bar{0}}}\\
\end{pmatrix}
\end{equation}
where

\begin{equation}{\label{eq:permittivity_elements}}
\begin{cases}
  \bar{\bar{\boldsymbol\epsilon}}_{\perp\perp}=
    \begin{pmatrix}
    \epsilon_{xx} & \epsilon_{xy}\\
    \epsilon_{yx} & \epsilon_{yy}\\
    \end{pmatrix},
  \quad\bar{\boldsymbol\epsilon}_{\perp}=
  \begin{pmatrix}
    \epsilon_{xz} \\
    \epsilon_{yz} \\
    \end{pmatrix},\\
  \quad\bar{\boldsymbol\epsilon}_{z}=
  \begin{pmatrix}
    \epsilon_{zx} & \epsilon_{zy}\\
  \end{pmatrix},
  \mathbf{\bar{\bar{I}}}_2=
  \begin{pmatrix}
    1 & 0\\
    0 & 1\\
  \end{pmatrix}
\end{cases}
\end{equation}

\subsection{M tensor eigenvalues and eigenvectors}
\par
Following \autoref{eq:wave_eq}, we need to investigate the eigenvalues and eigenvectors of the $\mathbf{M}$ tensor (\autoref{eq:Melements}) in order to calculate the tangential components of the electric and magnetic field. The eigenvectors $\mathbf{u}_i$~and $\mathbf{v}_i$ correspond to the eigenvalues $+\lambda_i$ and $-\lambda_i$ ($i=1,2$) for the forward and backward waves, respectively:
\begin{equation}{\label{eq:Eigenvectors}}
\begin{cases}
\mathbf{u}_i=\{E_{ix},E_{iy},-\eta_0H_{iy},\eta_0H_{ix}\}   \\
\quad\quad\quad\quad\quad\quad\quad\quad\quad\quad\quad\quad\quad\quad\quad\quad\text{for i=1,2}\\
\mathbf{v}_i=\{-E_{ix},-E_{iy},-\eta_0H_{iy},\eta_0H_{ix}\}
\end{cases}
\end{equation}

For a full analytical derivation of the eigenvalues and eigenvectors see section \textbf{``Theoretical analysis of the propagation matrix $\mathbf{M}$''}.

\subsection{E-field evolution in the propagation direction}
\par
To understand the \textbf{E}-field and \textbf{H}-field properties while the wave is propagating in the $z$ direction, we can employ the fundamental equation for one-dimensional wave propagation (\autoref{eq:wave_eq}). By considering the eigenvectors and eigenvalues of the $\mathbf{M}$ matrix (\autoref{eq:Eigenvectors}), the wave properties along the propagation direction ($z$) is calculated for the forward wave:
\begin{equation}{\label{eq:Z_direction1}}
\frac{d}{dz}\mathbf{u}_i=ik_0\lambda_i\mathbf{u}_i \Longrightarrow \mathbf{u}_i(z)=\mathbf{u}_i(0)e^{ik_0\lambda_iz}\\
\end{equation}

It is possible to calculate the backward wave, $\mathbf{v}_i(z)$, in the same way.

\par
By convention, the polarization of light is described by specifying the orientation of the wave's electric field at a point in space over one period of the oscillation. Therefore, to understand the polarization direction inside the medium it is enough to only derive the \textbf{E}-field. The initial light polarization before interaction with the medium could be described by the \textbf{E}-field:
\begin{eqnarray}{\label{eq:E_initial}}
\mathbf{E}_0=E_{0x}\mathbf{\hat{x}}+E_{0y}\mathbf{\hat{y}}
\end{eqnarray}
In addition, the transverse electric field ($x$-$y$ components) inside the medium could be written as a sum of the \textbf{E}-field of the two eigenvectors:
\begin{eqnarray}{\label{eq:E_inside}}
\mathbf{E}_{inside}&=&A\mathbf{E}_1(z)+B\mathbf{E}_2(z)\nonumber\\
          &=&A(E_{1x}\mathbf{\hat{x}}+E_{1y}\mathbf{\hat{y}})+B(E_{2x}\mathbf{\hat{x}}+E_{2y}\mathbf{\hat{y}})
\end{eqnarray}

Based on the boundary condition for the tangential E-fields at the interface:
\begin{equation}{\label{eq:BC}}
\begin{cases}
\mathbf{\hat{x}}:E_{0x}=A E_{1x}+B E_{2x}\\
\mathbf{\hat{y}}:E_{0y}=A E_{1y}+B E_{2y}\\
\end{cases}
\end{equation}
Therefore
\begin{equation}{\label{eq:alpha_beta}}
\begin{cases}
A=\frac{E_{0x}E_{2y}-E_{0y}E_{2x}}{E_{1x}E_{2y}-E_{1y}E_{2x}}\\
B=\frac{1}{E_{2x}}[E_{0x}-A E_{1x}]
\end{cases}
\end{equation}
\par

Now we can extend our fields to a different $z$ using \autoref{eq:Z_direction1}, \autoref{eq:E_inside} and \autoref{eq:alpha_beta} as follows:
\begin{equation}{\label{eq:Z_direction2}}
\mathbf{u}(z)=A \mathbf{u}_1e^{ik_0\lambda_1z}+B\mathbf{u}_2e^{ik_0\lambda_2z}\\
\end{equation}

Note that since the $\mathbf{M}$ matrix does not depend on $z$ in this case the eigenvalues and eigenvectors $+\lambda_i$, $-\lambda_i$, $\mathbf{u}_i$ and $\mathbf{v}_i$ remains constant and it is therefore possible to propagate the fields (or equivalently $\mathbf{u}(z)$ in \autoref{eq:Z_direction2}) to any desired length in one step. In the more general case where the eigenvalues and eigenvectors changes with $z$ one has to recalculate these in the simulations every distance $\Delta z$ to be able to propagate the fields to $z$, where $\Delta z$ is chosen for numerical stability and resolution.

Note also that the two polarization steady states for a forward propagating wave discussed in the main text is simply $\mathbf{u}_1$ and $\mathbf{u}_2$ where the solution with the lowest imaginary part of its eigenvalue is the stable solution whilst the other is the unstable solution. This can be understood since a high absorption leads to a fast decay from that state, i.e. the highest absorption gives the unstable state. In the same way, a large negative imaginary part corresponding to a high gain will generate a large component and therefore become stable since the other component can be neglected.

\subsection{Light propagation in a \pryso ~crystal}
\par
We start by calculating the permittivity tensor for \pryso ~and then employ the theoretical approach discussed above to obtain the light polarization direction while propagating through this specific medium.

\par
To define the permittivity tensor for \pryso ~one could start by deriving the relation between permittivity and susceptibility for the host material (\yso) and the absorber (Pr) separately as follows:
\begin{equation}{\label{eq:permittivity}}
\boldsymbol\epsilon =(1+\boldsymbol\chi)=(1+\boldsymbol\chi_{host}+i\boldsymbol\chi_{abs})
\end{equation}
The imaginary part of the susceptibility is proportional to the absorption, while the real part of the susceptibility is proportional to the real refractive index.

In section \textbf{``Theoretical analysis of the propagation matrix $\mathbf{M}$''} a full analytical derivation of the electric susceptibility $\chi_{abs}$ and the transition dipole moment angle $\theta$ is given but the following equations (\autoref{eq:suscept_permitt} and \autoref{eq:chi_absorption}) are approximately correct for the case of \pryso.

$\chi_{abs}$ is a dimensionless quantity which represents the imaginary part of the electric susceptibility and is approximately proportional to the absorption coefficients ($\alpha_{D_1}+\alpha_{D_2}$), in the medium as follows \cite{Haug2004}:
\begin{equation}{\label{eq:suscept_permitt}}
\chi_{abs}=\frac{n_{bg}c_0}{\omega}(\alpha_{D_1}+\alpha_{D_2})
\end{equation}
where $n_{bg}\approx1.8$ is a background refractive index and $\omega\approx2 \pi \cdot 494\cdot 10^{12}$ rad/s.
\par
The anisotropic absorption coefficients for the two sites of a nominally $0.05\%$ \pryso ~crystal at the \dhdh transition is measured as shown in the Table 7.6 in Ref. \cite{GuokuiLiu2005}. Based on this measurement the absorption for site I (605.977 nm) is about $3.6\pm0.5$ along the $D_1$, $47\pm5$ along the $D_2$, and $<0.1$ along the $b$ axis, all in the unit $\text{cm}^{-1}$ (see \autoref{fig:dipoleMoment}a, where $D_1$, $D_2$ and $b$ are crystal principal axes). To estimate the susceptibility and the transition dipole moment angle $\theta$, \autoref{eq:suscept_permitt} and the following equation are used (for more correct values see \autoref{eq:chi_absorption2});
\begin{equation}{\label{eq:chi_absorption}}
\begin{cases}
	\text{tan}(\theta)^2 &= \frac{\alpha_{D_2}}{\alpha_{D_1}}		\\
	\Rightarrow \theta &= 74.5 \pm 1.9\degree\\
	\Rightarrow \chi_{abs} &= (8.8\pm 1.0)\cdot10^{-4}
\end{cases}
\end{equation}
where $\chi_{abs}$ is given in the direction of the transition dipole moment shown in \autoref{fig:dipoleMoment}a. It can be rotated about the $b$ axis to the principal axis coordinate system to be able to perform the summation in \autoref{eq:permittivity};
\begin{equation}{\label{eq:rotation}}
\begin{split}
\boldsymbol\chi_{abs}^{\text{crystal}} &= \mathbf{R_z}^{-1}\cdot\boldsymbol\chi_{abs}^{\text{dipole}} \cdot \mathbf{R_z} \\
& =\chi_{abs}
\begin{pmatrix}
  \cos^2\theta & \sin\theta\cos\theta & 0\\
  \sin\theta\cos\theta & \sin^2\theta & 0\\
  0 & 0 & 0\\
\end{pmatrix}
\end{split}
\end{equation}

The next step is to calculate $\boldsymbol\chi_{host}$ based on the Sellmeier dispersion and the measured coefficients from Table 7.1 and Eq. 7.2 in Ref. \cite{GuokuiLiu2005}.
\begin{equation}{\label{eq:kai_host}}
\boldsymbol\chi_{host}=
\begin{pmatrix}
  2.1973 & 0 & 0\\
  0 & 2.2726 & 0\\
  0 & 0 & 2.1867\\
\end{pmatrix}
\end{equation}

Adding up all calculations, the final results for the permittivity tensor in the principal axis coordinate will be:
\begin{equation}{\label{eq:permittivity_final1}}
\boldsymbol\epsilon=
\begin{pmatrix}
  \epsilon_{xx} & \epsilon_{xy} & \epsilon_{xz}\\
  \epsilon_{yx} & \epsilon_{yy} & \epsilon_{yz}\\
  \epsilon_{zx} & \epsilon_{zy} & \epsilon_{zz}\\
\end{pmatrix}
\end{equation}
\begin{equation}{\label{eq:permittivity_final2}}
=
\begin{pmatrix}
  1+\chi_{xx}+i\chi_{abs}\cos^2\theta & i\chi_{abs}\sin\theta\cos\theta     & 0\\
  i\chi_{abs}\sin\theta\cos\theta     & 1+\chi_{yy}+i\chi_{abs}\sin^2\theta & 0\\
  0                                   & 0                                   & 1+\chi_{zz}\\
\end{pmatrix}\\
\end{equation}
\begin{equation}{\label{eq:permittivity_final3}}
=
\begin{pmatrix}
  3.1973+6.2e^{-5}i & 2.3e^{-4}i      & 0\\
  2.3e^{-4}i       & 3.2726+8.2e^{-4}i & 0\\
  0                  & 0                 & 3.1867\\
\end{pmatrix}
\end{equation}

Using this permittivity tensor together with \autoref{eq:Melements}, \autoref{eq:Eigenvectors}, \autoref{eq:alpha_beta} and \autoref{eq:Z_direction2} it is possible to propagate light through an absorbing \pryso ~crystal.\\
\par
\section{Theoretical analysis of the propagation matrix $\mathbf{M}$}
For light propagating parallel to one crystal axis of a plane-stratified, non-magnetic birefringent crystal with an absorption (or gain) axis in the transverse $x$-$y$ plane and no crossed magneto-electrical properties the propagation matrix $\mathbf{M}$ can be written as;

\begin{equation}{\label{eq:M}}
\mathbf{M}=
\begin{pmatrix}
  0 & 0 & -1 &  0 \\
  0 & 0 &  0 & -1 \\
  -n_{D_1}^2 - i\chi_{abs}\cos^2 \theta & -i\chi_{abs}\cos \theta \sin \theta & 0 & 0 \\
 -i\chi_{abs}\cos \theta \sin \theta & -n_{D_2}^2 - i\chi_{abs}\sin^2 \theta & 0 & 0 \\
\end{pmatrix}
\end{equation}
where $n_{D_1}$ and $n_{D_2}$ are the refractive indices of the two crystal axes in the $x$-$y$ plane. $\theta$ is the angle from $D_1$ to the transition dipole moment axis and $\chi_{abs}$ is the electric susceptibility associated with the absorption.

The polarization steady states are given by the eigenvectors to the propagation matrix $\mathbf{M}$. Four eigenvectors and eigenvalues exists and two of the eigenvalues have a positive real part which means they are connected to a forward traveling wave. These eigenvalues and eigenvectors are most important for the analysis in this paper. The stable eigenvector is the one whose eigenvalue has the lowest imaginary part (which is true for both positive imaginary parts in the case of absorption and for negative imaginary part in the case of gain).

\begin{widetext}
Solving for the eigenvalues of $\mathbf{M}$ gives four eigenvalues;

\begin{equation}{\label{eq:eigenvalues}}
\lambda_i = \pm \left( \frac{n_{D_1}^2 + n_{D_2}^2 + i \chi_{abs}}{2} \mp \frac{\chi_{abs}}{2}\sqrt{  \left(\frac{n_{D_2}^2 - n_{D_1}^2}{\chi_{abs}}\right)^2 - 2i(1-2\sin^2\theta)\frac{n_{D_2}^2 - n_{D_1}^2}{\chi_{abs}} -1}\right)^{1/2}
\end{equation}

\end{widetext}

Since we are only interested in the forward propagating waves only the eigenvalues whose real part is positive are of interest. These eigenvalues are denoted $\lambda_1$ and $\lambda_2$.

Before we solve for the eigenvectors in the general case we have to account for some special cases where $\theta = 0\degree$ or $\theta = 90\degree$. In these cases the transition dipole moment axis is parallel to one of the crystal axes $D_1$ or $D_2$ which of course results in two steady states, one along $D_1$ and another along $D_2$. Another special case is when the absorption is zero, which again is a trivial case where the polarizations of the steady states lie along $D_1$ and $D_2$.

Now that these special cases have been solved we will solve for the eigenvectors in the general case where we exclude the special cases mentioned above. The forward eigenvectors will be called $\mathbf{u}_i$;

\begin{equation}{\label{eq:Eigenvectors2}}
\mathbf{u}_i=\{E_{ix},E_{iy},-\eta_0H_{iy},\eta_0H_{ix}\}   \\
\end{equation}

and the propagation is given by;

\begin{equation}{\label{eq:u_prop}}
\begin{cases}
	\mathbf{u}_1(z)=\mathbf{u}_{1}(0)e^{ik_0\lambda_1z} \\
	\mathbf{u}_2(z)=\mathbf{u}_{2}(0)e^{ik_0\lambda_2z}
\end{cases}
\end{equation}

\begin{widetext}
The eigenvector equations we want to solve are;

\begin{equation}{\label{eq:eigenvectors_eq}}
\begin{pmatrix}
  0 & 0 & -1 &  0 \\
  0 & 0 &  0 & -1 \\
  -n_{D_1}^2 - i\chi_{abs}\cos^2 \theta & -i\chi_{abs}\cos \theta \sin \theta & 0 & 0 \\
 -i\chi_{abs}\cos \theta \sin \theta & -n_{D_2}^2 - i\chi_{abs}\sin^2 \theta & 0 & 0 \\
\end{pmatrix}
\begin{pmatrix}
  E_{ix} \\
  E_{iy} \\
  -\eta_0H_{iy} \\
   \eta_0H_{ix} \\
\end{pmatrix}
 = \lambda_i
\begin{pmatrix}
  E_{ix} \\
  E_{iy} \\
  -\eta_0H_{iy} \\
   \eta_0H_{ix} \\
\end{pmatrix}
\end{equation}

\end{widetext}

The first two equations provides a relationship between the electric and magnetic fields and are omitted here since we are only interested in the ratio between the $D_1$ and $D_2$ components of the electric field to get the steady state solution, i.e. we only have to solve for the first two components of the eigenvector. This leaves us with two equations;

\begin{equation}{\label{eq:eigenvectors_eq2}}
\begin{cases}
	(n_{D_1}^2 + i\chi_{abs}\cos^2 \theta)E_{ix} + (i\chi_{abs}\cos \theta \sin \theta) E_{iy}	= \lambda_i^2	E_{ix}	 \\
	( i\chi_{abs}\cos \theta \sin \theta)E_{ix} + (n_{D_2}^2 + i\chi_{abs}\sin^2 \theta)E_{iy}	= \lambda_i^2	E_{iy}
\end{cases}
\end{equation}

Since we have the option to normalize the eigenvector in any way we set $E_{iy} = 1$ (assuming that $E_{iy} \neq 0$). Now solving for the ratio $E_{ix} / E_{iy}$ we get;

\begin{equation}{\label{eq:ratio_electric}}
	E_{ix} = \frac{E_{ix}}{E_{iy}} = \frac{i\chi_{abs}\cos \theta \sin \theta}{\lambda_i^2-(n_{D_1}^2 + i\chi_{abs}\cos^2 \theta)}
\end{equation}

Using the following ratio;

\begin{equation} {\label{eq:ratio2}}
	R = \frac{n_{D_2}^2-n_{D_1}^2}{\chi_{abs}}
\end{equation}

\autoref{eq:ratio_electric} (with $\lambda_i$ from \autoref{eq:eigenvalues}) can be rewritten in the following way;

\begin{widetext}
\begin{equation}{\label{eq:ratio_electric2}}
	\frac{E_{ix}}{E_{iy}} = \frac{2\cos \theta \sin \theta}{1 - 2\cos^2\theta - iR \mp \sqrt{1 - (1-2\sin^2\theta)^2  +  (iR + (1-2\sin^2\theta))^2}}
\end{equation}
\end{widetext}

It can be shown that $\frac{E_{1x}}{E_{1y}} = -\frac{E_{2y}}{E_{2x}}$ which means that the two steady state solutions only differ by a $90\degree$ rotation in the $D_1$-$D_2$ plane since the two dimensional rotation matrix for an angle of $90\degree$ is

\begin{equation}
\mathbf{J} =
\begin{pmatrix}
  0 & -1 \\
  1 & 0 \\
\end{pmatrix}
\end{equation}

\subsection{Analysis when $\mathbf{R \rightarrow 0}$}

We can now solve for the steady state solutions (i.e. electric field components of the eigenvectors $\mathbf{u}_i$) in the limit $R \rightarrow 0$, which in other words means that the absorption is high compared to the phase retardation (see \autoref{eq:numerator_ratio}). The eigenvalues becomes;
\begin{equation}{\label{eq:eigenvalues_R0}}
\begin{cases}
	\lambda_1 \approx \sqrt{\frac{n_{D_1}^2 + n_{D_2}^2}{2}}\\
	\lambda_2 \approx \sqrt{\frac{n_{D_1}^2 + n_{D_2}^2}{2} + i\chi_{abs}}
\end{cases}
\end{equation}
i.e. one solution with no absorption (or gain) and a refractive index close to the refractive index of the material and one solution with high absorption (or gain) and a high refractive index.

The solution for the ratio of the electric field components becomes;

\begin{equation}{\label{eq:eigenvectors_R0}}
\begin{cases}
	\lambda_1: &\frac{E_{1x}}{E_{1y}} \approx -\tan(\theta)		\\
	\lambda_2: &\frac{E_{2x}}{E_{2y}} \approx \frac{1}{\tan \theta}
\end{cases}
\end{equation}

Since these are real numbers the polarizations are linear and the directions are for $\lambda_2$ along the transition dipole moment and for $\lambda_1$ perpendicular to it. Since the stable solution is the eigenvector whose eigenvalue has the lowest imaginary part, $\lambda_1$ is stable for absorption and $\lambda_2$ is stable for gain.

\subsection{Analysis when $\mathbf{R \rightarrow \infty}$}

In the other limit where $R \rightarrow \infty$ the phase retardation dominates over the absorption which gives the following approximation of the eigenvalues;

\begin{equation}{\label{eq:eigenvalues_Rinf}}
\begin{cases}
	\lambda_1 \approx \sqrt{n_{D_1}^2 +i\chi_{abs}\cos^2\theta}  \\
	\lambda_2 \approx \sqrt{n_{D_2}^2 +i\chi_{abs}\sin^2\theta}
\end{cases}
\end{equation}

Here the stable solution depends on the transition dipole moment direction given by $\theta$. If $\theta < 45\degree$ $\lambda_2$ is stable while if $\theta > 45 \degree$ $\lambda_1$ is stable for absorption and the reverse is true for gain. Using \autoref{eq:eigenvectors_eq2} we can solve for the eigenvectors in the limit when $R \rightarrow \infty$.
\begin{equation}{\label{eq:eigenvectors_Rinf}}
\begin{cases}
	\lambda_1: &(n_{D_1}^2 + i\chi_{abs}\cos^2 \theta)E_{1x} + (i\chi_{abs}\cos \theta \sin \theta) E_{1y} 	\\
	&\approx (n_{D_1}^2 + i\chi_{abs}\cos^2 \theta)E_{1x}									 \\
	&\Rightarrow E_{1y} \approx 0														 \\
	\lambda_2: &( i\chi_{abs}\cos \theta \sin \theta)E_{2x} + (n_{D_2}^2 + i\chi_{abs}\sin^2 \theta)E_{2y} 	\\
	&\approx  (n_{D_2}^2 + i\chi_{abs}\sin^2 \theta)E_{2y}									 \\
	&\Rightarrow E_{2x} \approx 0
\end{cases}
\end{equation}

From these equations it is clear that the eigenvector corresponding to $\lambda_1$ has almost no $E_{1y}$ component and is therefore along $D_1$ and the reverse is true for $\lambda_2$.

\subsection{Analysis of \pryso}
\par
In $0.05\%$ \pryso ~$R$ is quite high so we can use the approximations made in the previous section for when $R\rightarrow \infty$. Continuing from \autoref{eq:eigenvalues_Rinf} we can write the imaginary part of $\lambda_i$ as;

\begin{equation}{\label{eq:eigenvalues_Rinf_Pr}}
\begin{cases}
	\text{Im}(\lambda_1) \approx \frac{1}{2} \sqrt{2\sqrt{n_{D_1}^4+\chi_{abs}^2\cos^4\theta} - 2n_{D_1}^2}  \\
	\text{Im}(\lambda_2) \approx \frac{1}{2} \sqrt{2\sqrt{n_{D_2}^4+\chi_{abs}^2\sin^4\theta} - 2n_{D_2}^2}
\end{cases}
\end{equation}

\begin{widetext}
From these two equations we can solve for the transition dipole moment angle $\theta$ and the electric susceptibility associated with the absorption, $\chi_{abs}$, to be;

\begin{equation}{\label{eq:eigenvalues_theta_chi_Pr}}
\begin{cases}
	\theta \approx \arctan \left( \left(  \frac{  \left(2\cdot(\text{Im}(\lambda_2))^2+n_{D_2}^2\right)^2 - n_{D_2}^4 }{ \left(2\cdot(\text{Im}(\lambda_1))^2+n_{D_1}^2\right)^2 - n_{D_1}^4 } \right)^{1/4} \right) \\
	\chi_{abs} \approx \sqrt{ \left(2\cdot(\text{Im}(\lambda_1))^2+n_{D_1}^2\right)^2 - n_{D_1}^4 } + \sqrt{\left(2\cdot(\text{Im}(\lambda_2))^2+n_{D_2}^2\right)^2 - n_{D_2}^4}
\end{cases}
\end{equation}

\end{widetext}

In \pryso ~the exponential decays of the intensity along $D_1$ and $D_2$ are measured and they are $\alpha_{D_1} = 3.6 \pm 0.5 \text{ cm}^{-1}$ and $\alpha_{D_2} = 47 \pm 5 \text{ cm}^{-1}$ respectively. Given that the eigenvectors calculated in \autoref{eq:eigenvectors_Rinf} are along $D_1$ and $D_2$ respectively one can, using \autoref{eq:u_prop}, write the following equation;

\begin{equation}{\label{eq:eigenvalues_abs}}
\begin{cases}
	\text{Im}(\lambda_1) = \frac{\alpha_{D_1}}{2k_0} \\
	\text{Im}(\lambda_2) = \frac{\alpha_{D_2}}{2k_0} \\
	k_0 = \frac{2\pi}{\lambda_0}
\end{cases}
\end{equation}
where the factor of $1/2$ comes from that we now deal with electric fields and not intensities. $k_0$ is the wave number and $\lambda_0$ is the wavelength of the incoming wave in vacuum (not to be confused with $\lambda_1$ and $\lambda_2$ which are the eigenvalues of $\mathbf{M}$). Using \autoref{eq:eigenvalues_theta_chi_Pr} and \autoref{eq:eigenvalues_abs} one can calculate the angle $\theta$ from $D_1$ to the transition dipole moment axis and the absorption susceptibility $\chi_{abs}$ for a \pryso ~crystal using $\lambda_0 = 605.977$ nm, $n_{D_1} = 1.7881$ and $n_{D_2} = 1.809$.
\begin{equation}{\label{eq:chi_absorption2}}
\begin{cases}
	\theta &= 74.6 \pm 1.9\degree \\
	\chi_{abs} &= (8.82 \pm 0.8) \cdot10^{-4}
\end{cases}
\end{equation}
An approximation of $\theta$ and $\chi_{abs}$ in \autoref{eq:eigenvalues_theta_chi_Pr} can be made by assuming that $\text{Im}(\lambda_i) \ll  n_{D_i}$ and that $n_{D_1} \approx n_{D_2} \approx n_{bg}$ (which is the case for \pryso). This results in;

\begin{equation}{\label{eq:theta_approx}}
\begin{split}
	\theta &\approx  \arctan \left(  \left(  \frac{(\text{Im}(\lambda_2))^2 n_{D_2}^2}{(\text{Im}(\lambda_1))^2 n_{D_1}^2}  \right)^{1/4}     \right)	 \\
	&\approx \arctan \left(   \sqrt{\frac{\text{Im}(\lambda_2)}{\text{Im}(\lambda_1)}}    \right) = \arctan \left(   \sqrt{\frac{\alpha_{D_2}}{\alpha_{D_1}}}    \right)
\end{split}
\end{equation}

and

\begin{equation}{\label{eq:chi_approx}}
\begin{split}
	\chi_{abs} &\approx 2\text{Im}(\lambda_1)n_{D_1} + 2\text{Im}(\lambda_2)n_{D_2}  \\
	 &=\frac{\alpha_{D_1}n_{D_1}}{k_0} + \frac{\alpha_{D_2}n_{D_2}}{k_0} \approx  \frac{(\alpha_{D_1} + \alpha_{D_2})n_{bg}}{k_0}
\end{split}
\end{equation}

\autoref{eq:theta_approx} and \autoref{eq:chi_approx} are only valid in the case where $R \rightarrow \infty$ but allows us to use \autoref{eq:suscept_permitt} and \autoref{eq:chi_absorption} for the case of \pryso ~which gives values quite close to the more correct values obtained in \autoref{eq:chi_absorption2}.

\subsection{Continuation of analysis when $\mathbf{R \rightarrow 0}$}

In the other case where $R \rightarrow 0$ we only have one axis with absorption (see \autoref{eq:eigenvalues_R0}) and therefore in \autoref{eq:eigenvalues_abs} we replace $\alpha_{D_2}$ by $\alpha_{dipole}$ that measures the absorption along the transition dipole moment axis for $\lambda_2$ whose eigenvector points in that direction and set $\text{Im}(\lambda_1)=0$. $\theta$ is now given by the angle from the crystal axis to the axis of which the absorption is measured. Assuming that $\chi_{abs} \gg n_{D_i}^2$ the following proportionality between $\chi_{abs}$ and $\alpha_{dipole}$ can also be calculated using \autoref{eq:eigenvalues_R0};
\begin{equation}
	\chi_{abs} \approx \frac{1}{2}\left(\frac{\alpha_{dipole}}{k_0}\right)^2
\end{equation}
This shows that the electric susceptibility $\chi_{abs}$ does not always scale linearly with the absorption coefficient $\alpha_{dipole}$.

\end{document}